# Highly Anisotropic and Robust Excitons in Monolayer Black Phosphorus

Xiaomu Wang[1], Aaron M. Jones[2], Kyle L. Seyler[2], Vy Tran[3], Yichen Jia[1], Huan Zhao[4], Han Wang[4], Li Yang[3], Xiaodong Xu[2*], and Fengnian Xia[1*]

[1]Department of Electrical Engineering, Yale University, New Haven, Connecticut 06511

[2]Department of Physics and Department of Materials Science and Engineering, University of Washington, Seattle, Washington 98195

[3]Department of Physics, Washington University, St. Louis, Missouri, 63130

[4]Ming Hsieh Department of Electrical Engineering, University of Southern California, Los Angeles, CA 90089

**Abstract.** Semi-metallic graphene and semiconducting monolayer transition metal dichalcogenides (TMDCs) are the two-dimensional (2D) materials most intensively studied in recent years[1-2]. Recently, black phosphorus emerged as a promising new 2D material due to its widely tunable and direct bandgap, high carrier mobility and remarkable in-plane anisotropic electrical, optical and phonon properties[3,4,5,6,7,8,9,10,11]. However, current progress is primarily limited to its thin-film form, and its unique properties at the truly 2D quantum confinement have yet to be demonstrated. Here, we reveal highly anisotropic and tightly bound excitons in monolayer black phosphorus using polarization-resolved photoluminescence measurements at room temperature. We show that regardless of the excitation laser polarization, the emitted light from the monolayer is linearly polarized along the light effective mass direction and centers around 1.3 eV, a clear signature of emission from highly anisotropic bright excitons. In addition, photoluminescence excitation spectroscopy suggests a quasiparticle bandgap of 2.2 eV, from which we estimate an exciton binding energy of around 0.9 eV, consistent with theoretical results based on first-principles. The experimental observation of highly anisotropic, bright excitons with exceedingly large binding energy not only opens avenues for the future explorations of many-electron effects in this unusual 2D material, but also suggests a promising future in optoelectronic devices such as on-chip infrared light sources.

[*]Email: fengnian.xia@yale.edu and xuxd@uw.edu

Since early 2014, the 2D material research community has witnessed the rise of its newest member — black phosphorus[3,4,5,6,7,8,9,10,11]. With excellent transport properties, a bulk bandgap in the infrared, and strong in-plane anisotropy, layered black phosphorus[12,13,14,15,16] has regained remarkable interest for potential applications in high-performance thin-film electronics, mid- and near-infrared optoelectronics, and for developing conceptually novel devices which utilize its anisotropic properties[4,6,11]. Monolayer black phosphorus is of particular scientific and technological significance. In the monolayer limit, the strong 2D quantum confinement and reduced screening enhances the Coulomb interaction between carriers, allowing tightly bound excitons to form[17,18]. Moreover, with the inclusion of its monolayer form, black phosphorus is expected to have a very large thickness-dependent bandgap tuning range from 0.3 to around 2 electron volts[10,11]. However, past research has predominantly focused on black phosphorus in its multi-layer and thin-film forms. Sporadic studies[5,19] on mono- and few-layer phosphorus have yet to reveal the most fundamental and intriguing properties of this material in its true 2D limit.

In this work, we carried out a detailed study of monolayer black phosphorus using polarization-resolved photoluminescence spectroscopy. Combined with first principles calculation, our study provides compelling experimental evidence for the bandgap values (both optical and quasi-particle) of monolayer black phosphorus and also suggests an exciton binding energy of ~0.9 eV. In addition, we also reveal the highly anisotropic nature of these robust excitons, which may be utilized to build conceptually new optoelectronic and electronic devices.

Black phosphorus has an orthorhombic lattice structure belonging to the $D_{2h}$ point group. Each phosphorus atom is covalently bonded with three adjacent atoms to form a puckered honeycomb network as shown in Figure 1a. Hence, black phosphorus has reduced symmetry compared to graphene's $D_{6h}$ point group, resulting in a highly asymmetric band structure and unusually strong in-plane anisotropy. In our experiments, monolayer black phosphorus samples were prepared by micro-mechanical exfoliation of bulk crystals onto 285 nm or 90 nm $SiO_2$ thermally grown on silicon wafers. Figure 1b shows an optical micrograph of a monolayer flake on 285 nm oxide on silicon. We identify the black phosphorus monolayers by combining atomic force microscopy (AFM) and Raman spectroscopy characterizations (see Supplementary Information I and Methods for details). The thickness of monolayer flakes is measured to be around 0.7 nm. This value is slightly larger than the nominal monolayer black phosphorus thickness of 0.53 nm but significantly less than the expected bilayer thickness of 1.06 nm, suggesting that the flake is indeed a monolayer.

As discussed in the Supplementary Information I, Raman spectroscopy also reveals the crystal orientations of monolayer black phosphorus[4, 19]. The Raman spectra were measured by exciting with a linearly-polarized green laser (532 nm) incident along the z-direction perpendicular to the x-y plane shown in Fig. 1a. The incident light polarizations in x-y plane are shown in the upper right portion of Fig. 1a. Under this experimental configuration, three peaks can be observed near 470, 440 and 365 $cm^{-1}$ due to the selection rules, which corresponds to the $A_g^2$, $B_{2g}$ and $A_g^1$ vibration modes, respectively[20,21]. Since the atomic motion associated with the $A_g^2$ mode occurs primarily along the armchair (x) direction[20,22], its Raman scattering intensity is strongest with the excitation laser polarization aligned along x-axis as shown in Figs. 1c and 1d, thus providing an effective method to determine the crystalline direction of monolayer black phosphorus. We found that its intensity oscillation fits well to a $\cos^2(\theta)$ function ($\theta$ denotes the angle between the x-axis and the polarization of the excitation laser) plus an offset, as shown in Fig. 1d.

The polarization-resolved photoluminescence (PL) experiments on monolayer black phosphorus were performed with an excitation energy of 2.33 eV (532 nm) and a spot size of ~1.5 μm in vacuum at room temperature (see Methods for details). The laser power was kept below 20 μW to avoid any damage to the samples. Figure 2a shows the polarization-resolved photoluminescence spectra of a typical monolayer sample. Excitation polarization and PL detection are selectively oriented along either the x- or y-axes, leading to a total of four different spectra. Regardless of the excitation or detection polarization, the emission spectrum shows a single peak with a FWHM of about 150 meV centered at 1.31 eV (947 nm). For the five monolayer black phosphorus samples measured, the emission peaks fall into a range of 1.31 ± 0.02 eV. The highest PL intensity occurs when both excitation and detection polarizations are aligned with the x-direction. Moreover, it is worth mentioning that the emitted light is always predominantly polarized along the x-direction. The emission intensity along the y-direction is consistently less than 3% of that along the x-direction, regardless of the excitation light polarization, as illustrated in Figure 2b.

This intriguing observation of strongly polarized PL emission is consistent with previous first-principles simulations of monolayer black phosphorus, in which excited states are dominated by anisotropic excitons due to the reduced symmetry and screening[10]. Figure 2c depicts the electron wave function of the bright ground state exciton of monolayer black phosphorus calculated based on the first-principles GW-Bethe-Salpeter-Equation (BSE) theory (see Supplementary Information II for details). Since the wave function is strongly extended along the x-direction, the observation of highly polarized emission indicates the excitonic nature of the observed photoluminescence. The origin of this strong anisotropic excitonic behavior can be traced back to the anisotropic band dispersion of monolayer black phosphorus. The carriers are more mobile in the dispersive band along the x-direction than in the nearly-flat band along the y-direction; the isotropic Coulomb interaction thus binds the carriers most strongly in y- direction. Regardless of the excitation laser polarization, the high energy photons (2.33 eV) used in the PL experiments will first generate free electrons and holes in monolayer black phosphorus, from which anisotropic excitons form, leading to highly polarized light emission. Since the PL always originates from these anisotropic excitons, the emitted light is strongly polarized along x-direction, regardless of the excitation polarization as shown in Fig. 2b. However, since the absorption of monolayer black phosphorus depends on the excitation light polarization, the PL intensity does vary as the excitation polarization changes. Although excitonic light emission has been observed in monolayer transition metal dichalcogenides[23 24 25 26 27 28 29 30], our results represent the first experimental demonstration of anisotropic excitons in 2D materials. Here the excitonic properties of monolayer black phosphorus resemble those of quasi-one-dimensional systems, such as carbon nanotubes[31 32 33], whereas monolayer black phosphorus is a truly two-dimensional system. More importantly, the existence of such highly anisotropic excitons constitutes the first direct experimental evidence of distinctively different carrier mobilities along x- and y-directions in monolayer black phosphorus.

To determine the exciton binding energy, we performed photoluminescence excitation spectroscopy (PLE). Here, both excitation and detection are x-axis polarized. We monitored the exciton PL while the excitation wavelength was swept from 653 to 477 nm (1.9 to 2.6 eV) with constant excitation power (see Methods for details). Figure 3a shows a 2D plot of PLE spectra as a function of both excitation and emission photon energies. The sporadic sharp spikes are due to the imperfect subtraction of the radiation background. Figure 3b plots the PLE spectra taken along the horizontal blue, cyan, and green dashed lines in Fig. 3a, corresponding to excitation photon energies of 2.58, 2.45, and 2.34 eV, respectively. Regardless of the excitation photon energy, the PL peaks at 1.32 eV as expected since the emission originates from the same bright excitons. However, the intensity of PL is the strongest when the excitation energy is at 2.45 eV, six times stronger than with an excitation energy of 2.15 eV. The peak PL intensity as a function of excitation energy is plotted in the inset of Fig. 3b, where the data are taken along the vertical white dashed line in Fig. 3a.

To understand the origin of the significantly enhanced PL at an excitation energy of 2.45 eV, we compared the experimental results with calculated absorption spectra of monolayer black phosphorus, as shown in the lower panel of Fig. 3c (see Supplementary Information II for details about first principles calculations). The red and blue lines correspond to the calculated absorption curves with and without including e-h interactions in suspended monolayer, respectively. Clearly, the experimentally measured ground state exciton energy (1.32 ± 0.02 eV) matches the theoretical result (1.4 eV) reasonably well. Although the calculation is performed on a suspended monolayer to minimize the consumption of computational resources, it is known that the existence of a dielectric substrate reduces the quasiparticle bandgap and exciton binding energy in a similar manner and the calculated exciton energy does not depend strongly on the substrates[29, 34]. Moreover, the calculation indicates that the quasi-particle bandgap is around 2.2 eV, consistent with previously reported theoretical ($G_1W_1$) value.

The line shape of the calculated quasi-particle absorption spectrum, blue curve in the lower panel of Fig. 3c, is consistent with the data in the inset of Fig. 3b. The absorption of monolayer black phosphorus abruptly increases around the quasi-particle band edge and then decays, leading to similar behavior in PL intensity. Here, the quasi-particle band edge is defined as the position at which the absorption increase rate is the largest. Under this definition, the quasi-particle bandgap of monolayer black phosphorus can be extracted to be 2.26 ± 0.1 eV from the inset of Fig. 3b, which agrees with the calculated value (2.2 eV in lower panel of Fig. 3c). However, two minor factors need to be clarified in the comparison of experimental and theoretical results. First, in first principles simulations of quasi-particle absorption, the interference effect introduced by the 290 nm thick oxide and silicon substrate has not been taken into account. Calculations based on multilayer film model[35] show that the introduction of the interference effect does up shift the band edge by approximately 60 meV. As a result, if the interference effect is taken into account, the quasi-particle bandgap estimated using experiments should be modified to 2.2 ± 0.1 eV, which matches the theoretical results perfectly and leads to an exciton binding energy of 0.88 ± 0.12 eV. Second, the numerical calculation is performed on a suspended monolayer. If the substrate effect is taken into account, the quasi-particle band gap will probably be reduced by 0.1 to 0.2 eV, and then the experimentally extracted quasi-particle bandgap is in fact slightly larger than the theoretical one. The upper panel of Fig. 3c schematically shows the experimental ground state exciton energy, quasi-particle bandgap, and the exciton binding energy, namely, the energy difference between the excitonic emission peak and the quasi-particle bandgap. Finally, the absorption above the quasi-particle band edge is very similar to that of typical one-dimensional systems such as carbon nanotubes where the absorption drops after peaking. This quasi-one-dimensional behavior is again a direct experimental evidence of the anisotropic band dispersion of monolayer black phosorus.

In summary, we report the first experimental observation of highly anisotropic excitons with an extraordinarily large binding energy (0.88 ± 0.12 eV) in monolayer black phosphorus. The optical and quasi-particle bandgaps of monolayer black phosphorus are revealed to be 1.31 ± 0.02 and 2.2 ± 0.1 eV, respectively, which agree well with first principles calculations. Unveiling its unusual quasi-one-dimensional excitonic features advances the understanding of black phosphorus in its 2D quantum confined limit while opening the door for potential applications in infrared polarized emitters and optical communication networks. Finally, the experimental demonstration of anisotropic excitons also provides direct experimental evidence of the highly anisotropic carrier mobility in monolayer black phosphorus, which may be utilized to construct advanced electronic devices and circuits.

**Methods**

**Sample Preparation**: The monolayer black phosphorus samples were prepared by the standard mechanical exfoliation technique from bulk black phosphorus crystals (Smart Element Inc.). The as-prepared samples were sequentially cleaned by acetone, methanol and isopropanol (about 1 minute for each step) to remove the tape residue. After that, the samples were post-baked at 180 degrees for 5 minutes to remove the solvent residue.

**Atomic Force Microscopy (AFM) and Raman Spectroscopy**: The thin flakes were first identified by the color contrast of microscope images. A Bruker Dimension FastScan system was then used to determine the layer thickness accurately, thus establishing the relationship between the flake thickness and color contrast. Raman spectroscopy was performed using a Horiba HR LabRaman 300 system. We used a microscope with a 100× objective lens to focus the laser spot to about 1.5 μm. The incident laser power was attenuated to below 200 μW to minimize the sample damage. Due to the possible degradation of sample quality, we used freshly exfoliated sample in both AFM and Raman measurements.

**Polarization-resolved Photoluminescence (PL)**: The PL experiments were performed in vacuum to minimize the possibility of sample degradation. The excitation laser is linearly polarized. On the detection side, we utilized a half-wave plate and linear polarizer to select the x- and y- components of the emitted light. For photoluminescence excitation measurements, the excitation source came from a Fianium SC480 equipped with a LLTF Contrast filter, with both incident and detection polarizations aligned to the crystallographic x-axis. In order to minimize the impact of the Raman characterization on our samples, in PL measurements we always chose fresh sample areas that did not undergo Raman measurements.

**Acknowledgements**

F.X. acknowledges financial support from the Office of Naval Research (N00014-14-1-0565). M.J., K.S., and X.X. are supported by DoE, BES, Materials Sciences and Engineering Division (DE-SC0008145 and DE-SC0012509). Facilities use was supported by YINQE and NSF MRSEC DMR 1119826.

**Author contributions**

F.X., X.W. and H.W. conceived the projects. X.W. prepared samples and did Raman measurements. A.M.J., K.L.S. and X.W. performed the PL measurements. V.T. and L.Y. did the DFT calculations. Y.J. and H.Z. helped with sample preparations. F.X. and X.W. wrote the paper with input from L.Y., X.X. and H.W. F.X. and X.X. supervised the project. All the authors discussed the results and commented on the manuscript.

**Additional information**

Supplementary information is available in the online version of the paper. Reprints and permission information is available online at www.nature.com/reprints. Correspondence and requests for materials should be addressed to F.X. and X.X.

**Competing financial interests**

The authors declare no competing financial interests.

**Figure Captions**
**Figure 1: Characterization of monolayer black phosphorus.** **(a)** A schematic structural view of monolayer black phosphorus, showing the crystal orientation. **(b)** Optical micrograph of an exfoliated monolayer black phosphorus flake on 290 nm $SiO_2$ on silicon substrate. The monolayer region is indicated by the grey dashed line. Scale bar (white): 6 μm. Inset: atomic force micrograph line scan showing a height of 0.7 nm for monolayers. **(c)** Polarization-resolved Raman scattering spectra of monolayer black phosphorus with 532-nm linearly polarized laser excitation. Raman spectra with excitation laser polarization along x-(grey), $30^o$ (red), $60^o$ (green), and y-(blue) directions as defined in (a) are shown. The light is incident along the z-direction (normal to x-y plane in a) and the power remains unchanged in polarization-resolved measurements. The peak positions of all three modes ($A_g^2$, $B_{2g}$, and $A_g^1$) do not change as the excitation laser polarization angle varies, however the intensity of $A_g^2$ mode strongly depends on the excitation laser polarization. **(d)** The intensity of $A_g^2$ mode as a function of the excitation laser polarization angle in x-y plane.

**Figure 2: Exciton photoluminescence (PL) with large in-plane anisotropy. (a)** Polarization-resolved PL spectra, revealing the excitonic nature of the emission from monolayer black phosphorus. The excitation 532-nm laser is linearly polarized along either x- (grey curves) or y- (blue curves) directions. On the detection side, a half-wave plate and linear polarizer selects x- or y- polarized components of the emitted light, leading to a total of 4 different combinations as shown. **(b)** PL peak intensity as a function of polarization detection angle for excitation laser polarized along x-(grey), 45° (magenta) and y-(blue) directions. The excitation laser power remains constant. The PL emission along y-direction is always less than 3% of that in the x- direction, regardless of the excitation light polarization angle. **(c)** A top view of the square of the electron wave-function of the ground state exciton in monolayer black phosphorus. Since the carriers are more mobile along the x-direction with low effective mass and the Coulomb interaction is isotropic, the exciton is anisotropic, forming striped patterns as shown. Scale bar: 1 nm.

**Figure 3: Large exciton binding energy revealed by photoluminescence excitation (PLE) spectroscopy. (a)** Photoluminescence excitation intensity map as a function of both excitation and emission photon energies. **(b)** Selected PL spectra along the three horizontal lines shown in (a), showing that the PL is the strongest when excitation photon energy is about 2.45 eV (505 nm). Inset: PL intensity at the emission peak of 1.32 eV as a function of excitation photon energy. The estimated quasi-particle band edge (before interference correction) is indicated by the shaded region. The data are taken along the vertical white dashed line in (a). **(c)** Upper: a schematic plot showing the measured ground state exciton energy (narrow red band) and the energy corresponding to the quasi-particle band edge (wide blue band). The width of the red and blue bands indicates the experimental uncertainty. Lower: Calculated optical absorption of monolayer black phosphorus with e-h interactions (excitonic absorption, red curve) and without e-h interactions (quasi-particle absorption, blue curve).

# References

[1] Geim, A. K. & Novoselov, K. S. The rise of graphene. *Nature Mater.* **6**, 183-191, (2007).

[2] Wang, Q. H., Kalantar-Zadeh, K., Kis, A., Coleman, J. N. & Strano, M. S. Electronics and optoelectronics of two-dimensional transition metal dichalcogenides. *Nature Nanotech.* **7**, 699-712, (2012).

[3] Li, L. *et al.* Black phosphorus field-effect transistors. *Nat. Nanotechnol.* **9**, 372-377, (2014).

[4] Xia, F., Wang, H. & Jia, Y. Rediscovering black phosphorus as an anisotropic layered material for optoelectronics and electronics. *Nature Commun.* **5**, 4458 (2014).

[5] Liu, H. *et al.* Phosphorene: an unexplored 2D semiconductor with a high hole mobility. *ACS Nano* **8**, 4033-4041, (2014).

[6] Churchill, H. O. H. & Jarillo-Herrero, P. Two-dimensional crystals: phosphorus joins the family. *Nature Nanotech.* **9**, 330-331, (2014).

[7] Koenig, S. P., Doganov, R. A., Schmidt, H., Castro Neto, A. H. & Özyilmaz, B. Electric field effect in ultrathin black phosphorus. *Appl. Phys. Lett.* **104**, 103106, (2014).

[8] Rodin, A. S., Carvalho, A. & Castro Neto, A. H. Strain-induced gap modification in black phosphorus. *Phys. Rev. Lett.* **112**, 176801 (2014).

[9] Buscema, M. *et al.* Fast and broadband photoresponse of few-Layer black phosphorus field-effect transistors. *Nano Lett.* **14**, 3347-3352, (2014).

[10] Tran, V., Soklaski, R., Liang, Y. & Yang, L. Layer-controlled band gap and anisotropic excitons in few-layer black phosphorus. *Phys. Rev. B* **89**, 235319 (2014).

[11] Qiao, J., Kong, X., Hu, Z.-X., Yang, F. & Ji, W. High-mobility transport anisotropy and linear dichroism in few-layer black phosphorus. *Nature Commun.* **5**, 4475 (2014).

[12] Keyes, R. W. The electrical properties of black phosphorus. *Phys. Rev.* **92**, 580-584 (1953).

[13] Warschauer, D. Electrical and optical properties of crystalline black phosphorus. *J. Appl. Phys.* **34**, 1853-1860 (1963).

[14] Jamieson, J. C. Crystal structures adopted by black phosphorus at high pressures. *Science* **139**, 1291-1292 (1963).

[15] Wittig, J. & Matthias, B. T. Superconducting Phosphorus. *Science* **160**, 994-995 (1968).

[16] Morita, A. Semiconducting black phosphorus. *Appl. Phys. A* **39**, 227-242 (1986).

[17] Koch, S. W., Kira, M., Khitrova, G. & Gibbs, H. M. Semiconductor excitons in new light. *Nature Mater.* **5**, 523-531 (2006).

[18] Scholes, G. D. & Rumbles, G. Excitons in nanoscale systems. *Nature Mater.* **5**, 683-696 (2006).

[19] Zhang, S. *et al.* Extraordinary photoluminescence and strong temperature/angle-dependent Raman responses in few-Layer phosphorene. *ACS Nano*, ASAP (2014).

[20] Sugai, S. & Shirotani, I. Raman and infrared reflection spectroscopy in black phosphorus. *Solid State Commun.* **53**, 753-755 (1985).

[21] Akahama, Y., Kobayashi, M. & Kawamura, H. Raman study of black phosphorus up to 13 GPa. *Solid State Commun.* **104**, 311-315 (1997).

[22] Fei, R. & Yang, L. Lattice vibrational modes and Raman scattering spectra of strained phospherene. *Appl. Phys. Lett.* **105**, 083120 (2014).

[23] Mak, K. F., Lee, C., Hone, J., Shan, J. & Heinz, T. F. Atomically thin $MoS_2$: a new direct-gap semiconductor. *Phys. Rev. Lett.* **105**, 136805 (2010).

[24] Splendiani, A. *et al.* Emerging Photoluminescence in Monolayer $MoS_2$. *Nano Lett.* **10**, 1271-1275 (2010).

[25] Eda, G. et al. Photoluminescence from chemically exfoliated $MoS_2$. *Nano Lett.* **11**, 5111-5116 (2011).

[26] Zeng, H., Dai, J., Yao, W., Xiao, D. & Cui, X. Valley polarization in $MoS_2$ monolayers by optical pumping. *Nature Nanotech.* **7**, 490-493 (2012).

[27] Mak, K. F. *et al.* Tightly bound trions in monolayer $MoS_2$. *Nature Mater.* **12**, 207-211 (2013).

[28] Ross, J. S. *et al.* Electrical control of neutral and charged excitons in a monolayer semiconductor. *Nature Commun.* **4**, 1474 (2013).

[29] Ugeda, M. M. *et al.* Giant bandgap renormalization and excitonic effects in a monolayer transition metal dichalcogenide semiconductor. *Nature Mater.* AOP doi:10.1038/nmat4061 (2014).

[30] Ye, Z. *et al.* Probing excitonic dark states in single-layer tungsten disulphide. *Nature* **513**, 214-218 (2014).

[31] Spataru, C. D., Ismail-Beigi, S., Capaz, R. B. & Louie, S. G. Theory and Ab initio aalculation of radiative lifetime of excitons in semiconducting carbon nanotubes. *Phys. Rev. Lett.* **95**, 247402 (2005).

[32] Wang, F., Dukovic, G., Brus, L. E. & Heinz, T. F. The optical resonances in carbon nanotubes arise from excitons. *Science* **308**, 838-841 (2005).

[33] Dresselhaus, M. S., Dresselhaus, G., Saito, R. & Jorio, A. Exciton photophysics of carbon nanotubes. *Annu. Rev. Phys. Chem.* **58**, 719-747 (2007).

[34] Yang, L. Excitonic effects on optical absorption spectra of doped graphene. *Nano Lett.* **11**, 3844-3847 (2011).

[35] Blake, P. *et al*. *Appl. Phys. Lett.* **91**, 063124 (2007).

# Figure 1: Characterization of monolayer black phosphorus

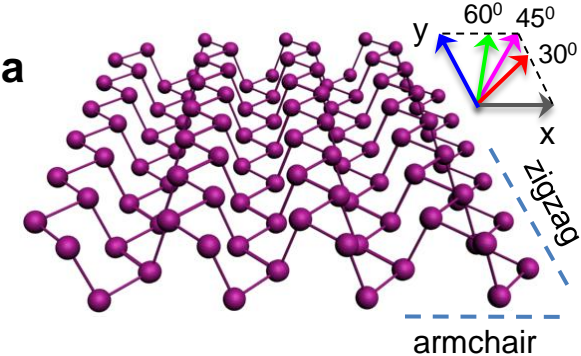
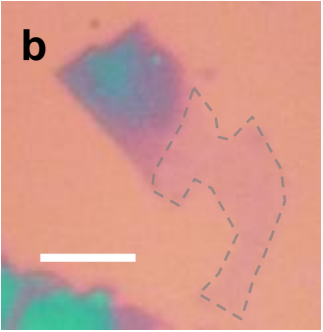
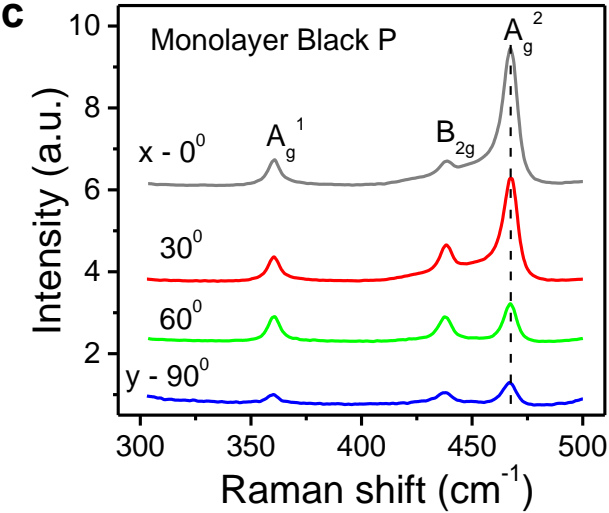
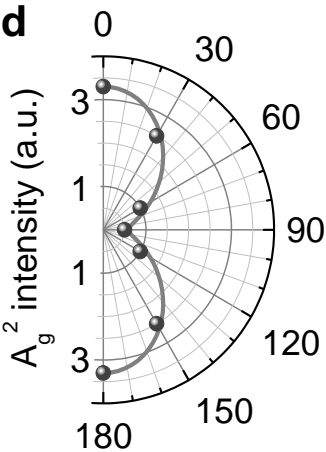

# Figure 2: Exciton photoluminescence (PL) with large in-plane anisotropy

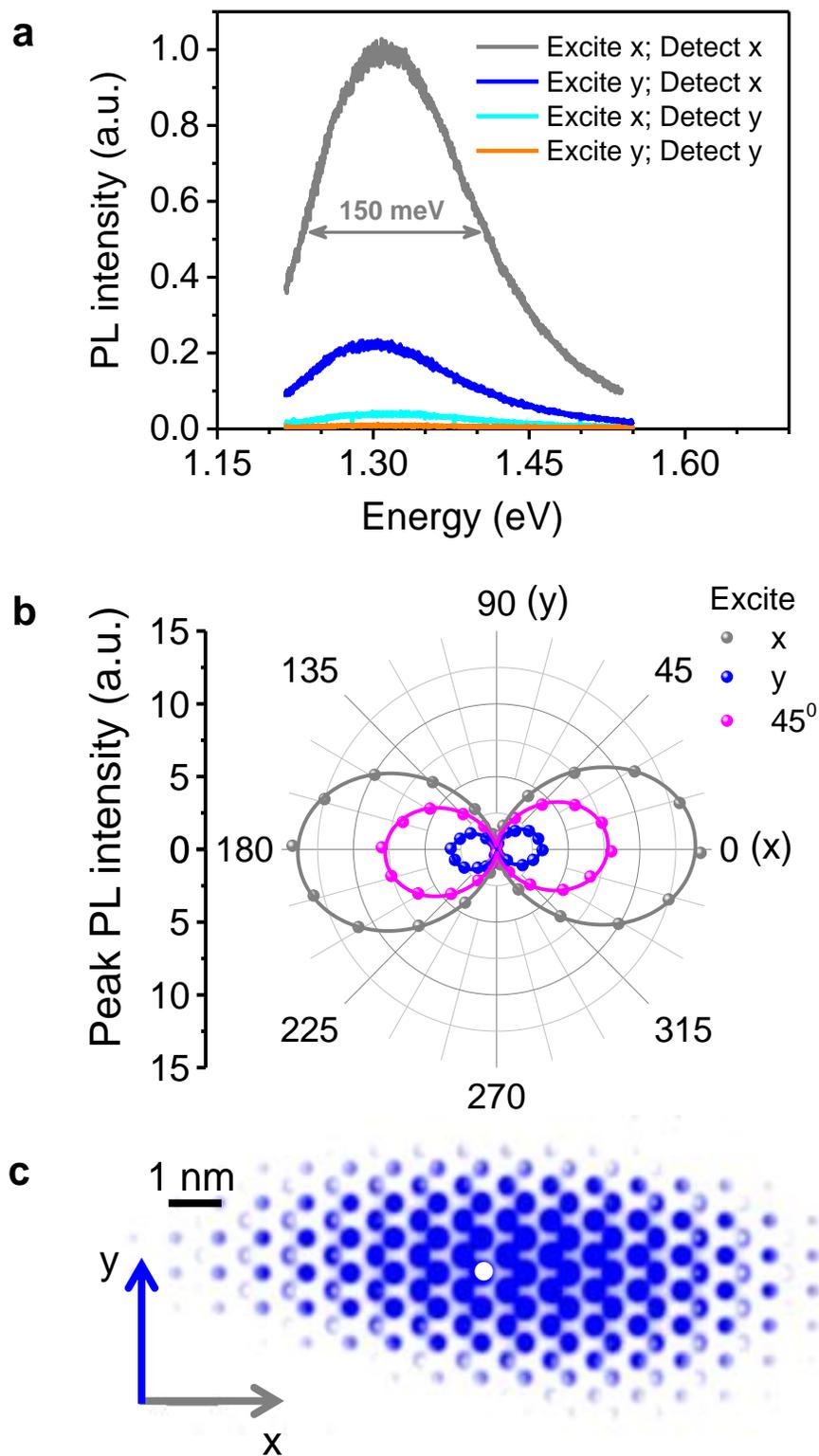

**Figure 3: Large exciton binding energy revealed by photoluminescence excitation (PLE) spectroscopy**

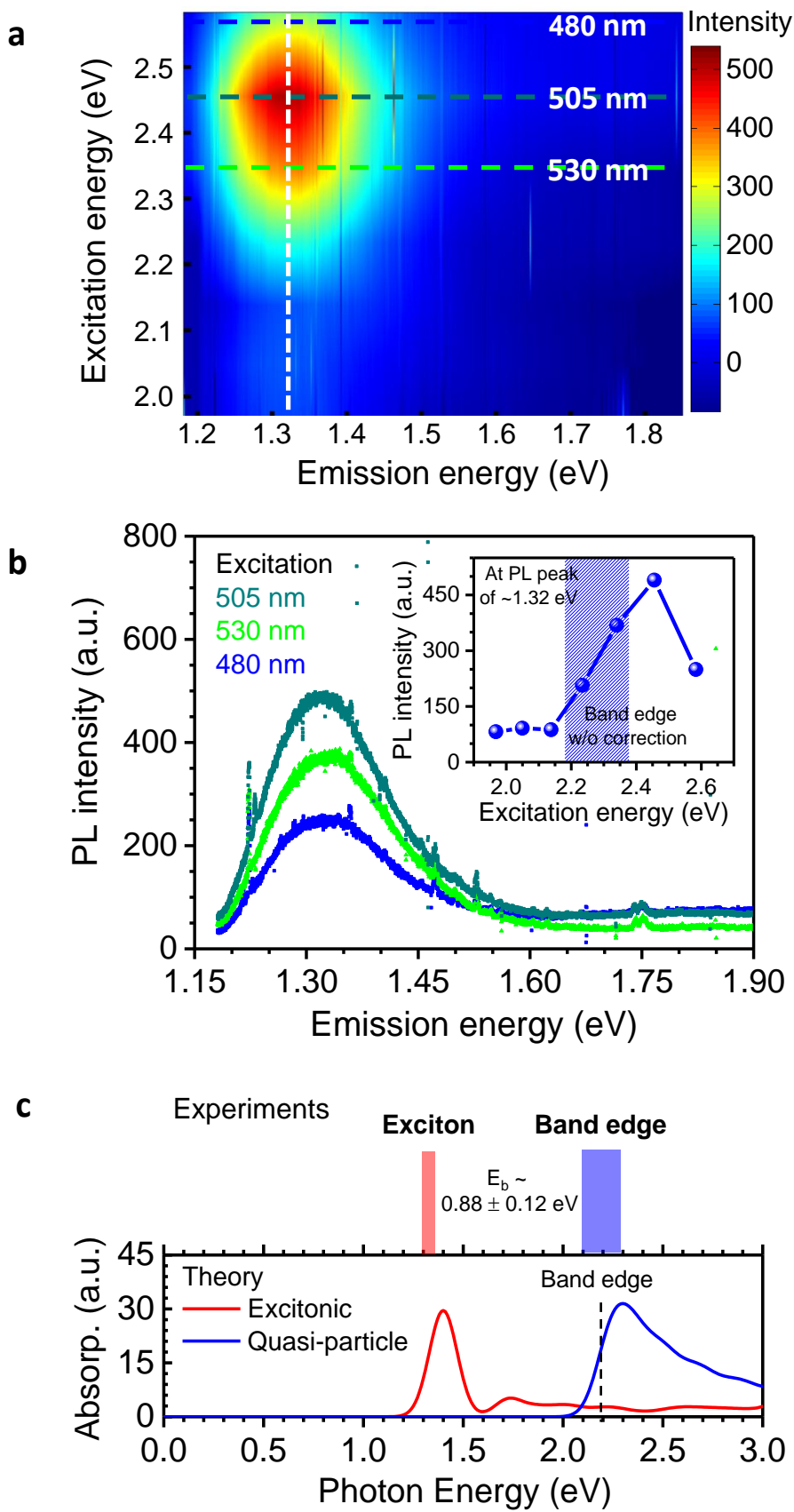

# Supplementary Information for "Highly Anisotropic and Robust Excitons in Monolayer Black Phosphorus"

Xiaomu Wang[1], Mitch Jones[2], Kyle Seyler[2], Vy Tran[3], Yichen Jia[1], Huan Zhao[4], Han Wang[4], Li Yang[3], Xiaodong Xu[2*], and Fengnian Xia[1*]

[1]Department of Electrical Engineering, Yale University, New Haven, Connecticut 06511

[2]Department of Physics and Department of Materials Science and Engineering, University of Washington, Seattle, Washington 98195

[3]Department of Physics, Washington University, St. Louis, Missouri, 63130

[4]Ming Hsieh Department of Electrical Engineering, University of Southern California, Los Angeles, CA 90089

*Email: fengnian.xia@yale.edu and xuxd@uw.edu

## I. Determination of Black Phosphorus Layer Number

In this work, we combine the atomic force microscopy (AFM) and Raman spectroscopy to identify monolayers and their crystalline orientation. We first located very thin, freshly exfoliated flakes visually and then measured their thicknesses using an atomic force microscope. The monolayer black phosphorus was found to be around 0.7 nm thick, well below the nominal thickness of two layers of black phosphorus (~ 1.1 nm). The bilayer thickness was measured to be within 1.1 to 1.6 nm. Figures S1a and b show the AFM images of fresh mono- and bilayer flakes, respectively, and the line scans showing the height are presented in the insets.

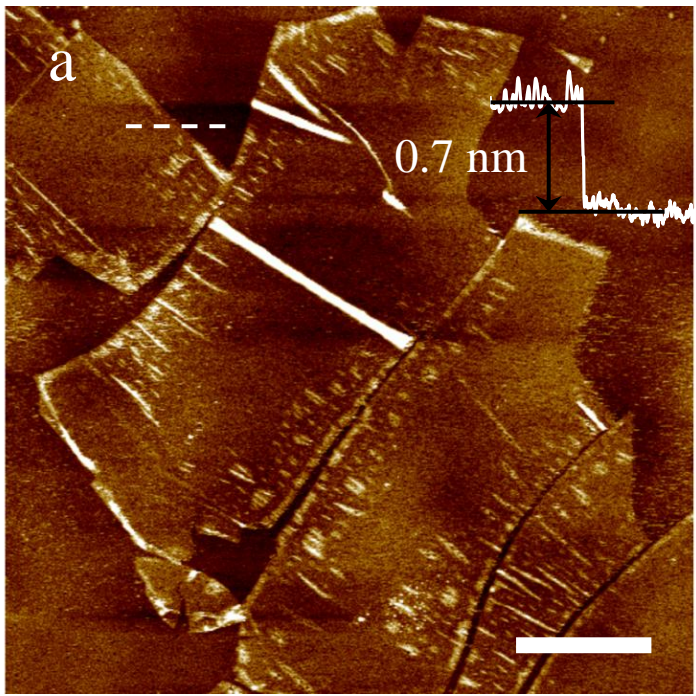

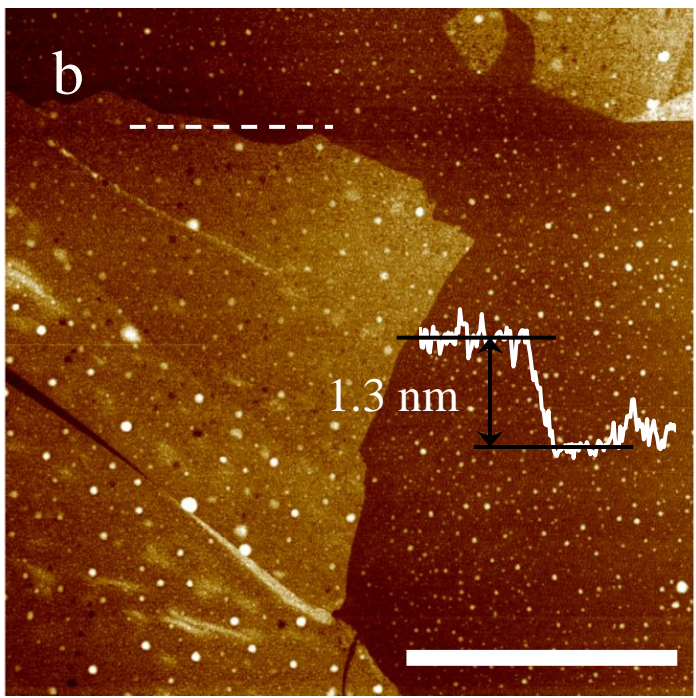

Figure S1. AFM images of black phosphorus. (a) Monolayer sample and (b) bilayer sample. Insets: line scans performed along the white dashed lines shown in the AFM image. Scale bar in both (a) and (b): 2 μm.

Following AFM characterization, we performed Raman scattering measurements (with a resolution of 0.25 cm$^{-1}$) on black phosphorus monolayer, bilayer, and thin film (> 20 layers). Regardless of the layer number, the $A_g^2$ mode intensity always peaks when the excitation laser polarization is aligned along x-axis of the crystalline structure, because this mode involves primarily the in-plane atomic motions along the x-axis [S1, S2]. By rotating the samples, we obtained the Raman spectra with maximum $A_g^2$ intensity for flakes with different thicknesses as shown in Fig. S2. All three spectra were calibrated using the prominent silicon peak at 520.9 cm$^{-1}$. We found that the position of $A_g^2$ mode of monolayer black phosphorus is located at around 468.3 cm$^{-1}$, ~1.2 cm$^{-1}$ higher than that in bilayer. While for black phosphorus thin film (> 20 layers), the $A_g^2$ mode peak position is located at around 465.9 cm$^{-1}$, ~ 2.5 cm$^{-1}$ lower than in monolayer. This blue shift of the in-plane $A_g^2$ mode with reduction in layer number is likely due to the long-range Coulombic interlayer interactions, similar to the evolvement of $E_{2g}^1$ mode in few-layer molybdenum disulphide ($MoS_2$) [S3]. Moreover, the full-width-half-maximum (FWHM) of the $A_g^2$ mode increases as the layer number goes down, probably due to the enhanced interaction with external environment. For black phosphorus thin film, the FWHM is around 5 cm$^{-1}$, and it increases to 7.4 cm$^{-1}$ and 8.5 cm$^{-1}$ for bilayer and monolayer, respectively. In addition, we also observed that the intensity ratio of $A_g^2$ to $A_g^1$ ($A_g^1$ represents the out-of-plane variations) is also layer-number dependent. For instance, the $A_g^2$ to $A_g^1$ intensity ratio is always greater than 5 for monolayer. The ratio is around 4 for bilayer and reduces to less than 3 for thin film (> 20 layers). After establishing the relationship between the layer number and Raman spectra, we were able to quickly identify the monolayers using Raman spectroscopy.

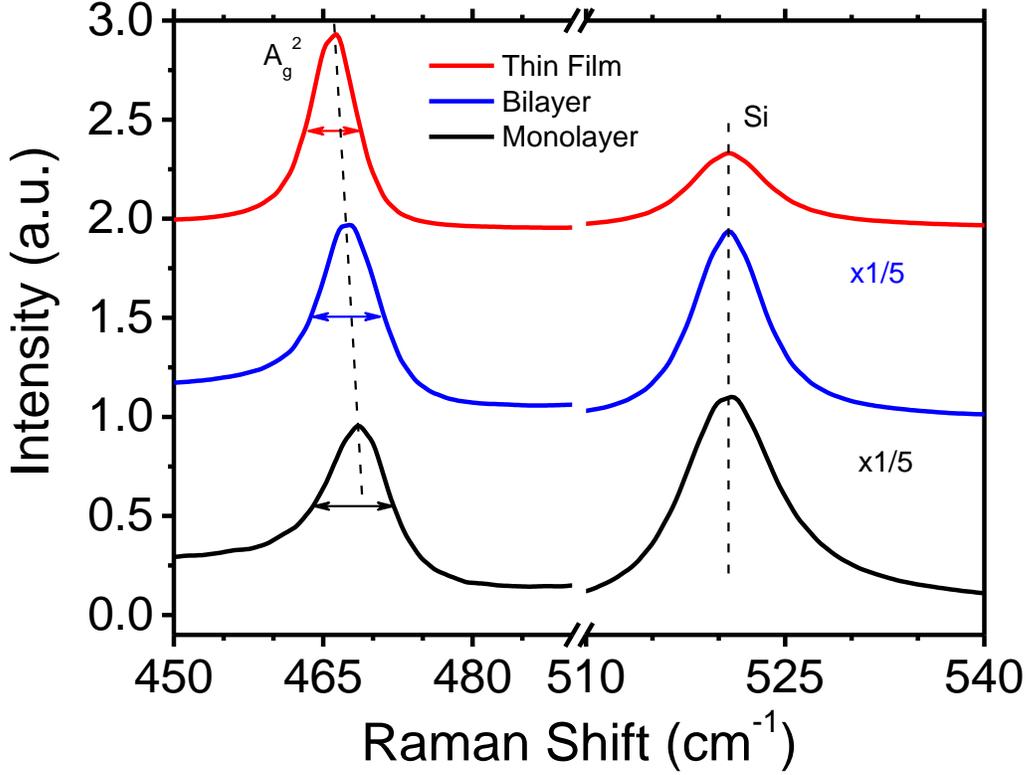

Figure S2. Raman spectra showing black phosphorus $A_g^2$ mode and silicon peak at different layer thickness. The excitation laser polarization is always aligned along the x-axis of crystalline structures.

## II. First Principles Calculations

We fully relax the atomic structures according to the force and stress calculated by density functional theory (DFT) within the Perdew, Burke and Ernzerhof (PBE) functional [S4]. All symmetries are slightly broken to mimic realistic samples. The ground-state wave functions and eigenvalues are calculated via DFT/PBE with a k-point grid of 14×10×1. We use a plane-wave basis with a 25 Ry energy cut-off with a norm-conserving pseudo-potential [S5]. The quasiparticle band gap is calculated by the GW approximation [S6] using the general plasmon pole model with a 21×15×1 k-point grid. The Green's function and dielectric screening are iterated one time to improve the convergence. The involved unoccupied band number is about 10 times that of the valence bands, producing a converged dielectric function. The excitonic effects are included by solving the Bethe-Salpeter Equation (BSE) with a finer k-point grid of 56×40×1 [S7]. A slab Coulomb truncation is employed to mimic the monolayer structure [S8, S9].

[S1] S. Sugai, S. and I. Shirotani. *Solid State Commun.* **53,** 753 (1985).

[S2] Y. Akahama, M. Kobayashi, and H. Kawamura. *Solid State Commun.* **104**, 311 (1997).

[S3] C. Lee, et al. *ACS Nano* **4**, 2695 (2010).

[S4] J. P. Perdew, K. Burke, and M. Ernzerhof. *Phys. Rev. Lett.* **77**, 3865 (1996).

[S5] N. Troullier and J. L. Martins. *Phys. Rev. B* **43**, 1993 (1991).

[S6] M. S. Hybertsen and S. G. Louie. *Phys. Rev. B* **34**, 5390 (1986).

[S7] M. Rohlfing and S. G. Louie. *Phys. Rev. B* **62**, 4927 (2000).

[S8] S. Ismail-Beigi. *Phys. Rev. B* **73**, 233103 (2006).

[S9] C. A. Rozzi, D. Varsano, A. Marini, E. K. U. Gross, and A. Rubio. *Phys. Rev. B* **73**, 205119 (2006).